\documentclass[conference]{IEEEtran}
\ifCLASSINFOpdf
\else
\fi
\usepackage[cmex10]{amsmath}
\usepackage{ amssymb }
\usepackage[pdftex]{graphicx}
\usepackage{float}
\usepackage{multicol}
\usepackage{lipsum}
\usepackage{ifpdf}
\usepackage{multirow}
\usepackage{epstopdf}
\usepackage{color}
\usepackage{cite}
\usepackage{graphicx}
\usepackage{soul}
\usepackage{balance}
\usepackage{threeparttable}
\usepackage{wasysym}
\usepackage[hyphens, spaces, obeyspaces]{url}
\usepackage[normalsize]{subfigure}
\usepackage{changepage} 
\usepackage{enumitem}
\usepackage{dblfloatfix}    
\usepackage{afterpage}
\usepackage{changepage} 
\usepackage[hidelinks, breaklinks]{hyperref}
\usepackage{accents}

\usepackage{algorithmicx}
\usepackage{algpseudocode}

\usepackage{fancyhdr}

\begin{document}
%
\IEEEpeerreviewmaketitle

\title{EMS and DMS Integration of the Coordinative Real-time Sub-Transmission Volt-Var Control Tool under High DER Penetration}
\vspace{-17ex}
%
%
%

\author{\IEEEauthorblockN{Quan Nguyen, Jim Ogle, Xiaoyuan Fan,\\
							Xinda Ke, Mallikarjuna R. Vallem, Nader Samaan\\
		\IEEEauthorblockA{Pacific Northwest National Laboratory, WA} \\[-7.0ex]
				\and
		\IEEEauthorblockN{Ning Lu}
		\IEEEauthorblockA{North Carolina State University, NC}
		\\[-7.0ex]
}
}

\maketitle


\vspace{-13ex}

\begin{abstract}
	This paper proposes an applicable approach to deploy the Coordinative Real-time Sub-Transmission Volt-Var Control Tool (CReST-VCT), and a holistic system integration framework considering both the energy management system (EMS) and distribution system management system (DMS). This provides an architectural basis and can serve as the implementation guideline of CReST-VCT and other advanced grid support tools, to co-optimize the operation benefits of distributed energy resources (DERs) and assets in both transmission and distribution networks. Potential communication protocols for different physical domains of a real application is included. Performance and security issues are also discussed, along with specific considerations for field deployment. Finally, the paper presents a viable pathway for CReST-VCT and other advanced grid support tools to be integrated in an open-source standardized-based platform that supports distribution utilities.
\end{abstract}

\begin{IEEEkeywords}
	Transmission system, Distribution system, Communication, Energy management system, Distribution management system, Co-optimization, Photovoltaic integration.
\end{IEEEkeywords}

\section{Introduction}
\IEEEPARstart{R}{enewable} 
	generation has been steadily increasing and reshaping the landscape of power generation. Increasing penetration of distributed energy resources (DERs) in distribution systems make them more and more dynamic, with different volt-var profiles from traditional centralized supply models. Therefore, industry, government organizations, and researchers have been developing methods and tools that can increase system operational efficiency and reliability: substation automation, automated fault location identification, volt/var control, etc. Recently, Eaton developed the CYME Volt/var Optimization analysis module to enable optimized distribution network operations \cite{Eaton_1}. In \cite{Etap_1}, ETAP introduced a volt/var optimization tool that optimally manages system-wide voltage levels and reactive power flow. Meanwhile, ABB, Siemens, Schweitzer Engineering Laboratory, and Dominion Energy developed volt/var optimization solutions to increase distribution system operational efficiency \cite{ABB_2, SEL_1}.
	
	With a sufficient penetration, DERs can have both positive and negative effects on the upstream transmission system. By controlling the power injected into or absorbed by the distribution systems, transmission system operators in the energy management system (EMS) control center can take advantage of the DERs and also mitigate their negative effects on the transmission system planning and operation. To achieve an optimal coordinative transmission and distribution (T\&D) operation, T\&D operators need to establish bidirectional communication and efficiently handle the computational effort required by the operation framework on each side. As the grid assets in transmission and distribution network are operated by different entities within a utility or across separate owners, these capabilities have yet to be implemented. Very limited information exchange and coordination exist between current EMS and distribution management system (DMS) control centers impedes the possibility of T\&D co-optimization.
		
	Research focusing on closing the gap between T\&D has recently been emerging. In \cite{Quan_4}, the	power flow model and solution of a combined T\&D system are discussed. In \cite{Zhengshuo_1}, a coordinated T\&D co-optimization framework is developed. In \cite{Zhao_1}, a coordinated restoration of a T\&D system using a decentralized scheme is proposed. As an extension of the existing works on co-optimization for integrated T\&D systems, the Coordinated Real-time Sub-Transmission Volt-var Control Tool (CReST-VCT) is proposed to coordinate and optimize the use of var resources in both T\&D sides  \cite{Xinda_1}. While these works mainly focus on the T\&D modeling and numerical analysis, aspects such as the communication and practical integration in EMS and DMS are not yet addressed.  
		
	In this paper, a high-level integration framework, practical challenges, and deployment approaches for a successful and reliable implementation of CReST-VCT across EMS and DMS are proposed and discussed. This work also serves as a  reference for implementing a T\&D analytical tool in real system control centers, and it can be tailored to a specific environment based on the corresponding constraints and requirements.
	
	The contribution of this paper includes:
	\begin{itemize}[leftmargin=*]
		\item A generalized design framework for CReST-VCT and other T\&D tools in EMS and DMS  for enabling high levels of DER penetration.
		\item An analysis of potential performance and security issues that should be considered during a practical implementation.
		\item An example of implementing the distribution part of CReST-VCT in an open-source platform used for designing, studying, and testing advanced operational grid functionalities. 
	\end{itemize}

\section{Brief Introduction of CReST-VCT}
	\subsection{Objective}
	CReST-VCT is an interactive tool that coordinates and optimizes the action of static and dynamic reactive power components, such as capacitor banks and PV inverters, respectively, from a sub-transmission system down to its distribution feeders \cite{Xinda_1}. CReST-VCT aims to enable high PV penetration at  the distribution feeders while alleviating or eliminating voltage violations in the sub-transmission system.
	\begin{figure}[b!]
		\centering
		\includegraphics[width = 0.85\columnwidth] {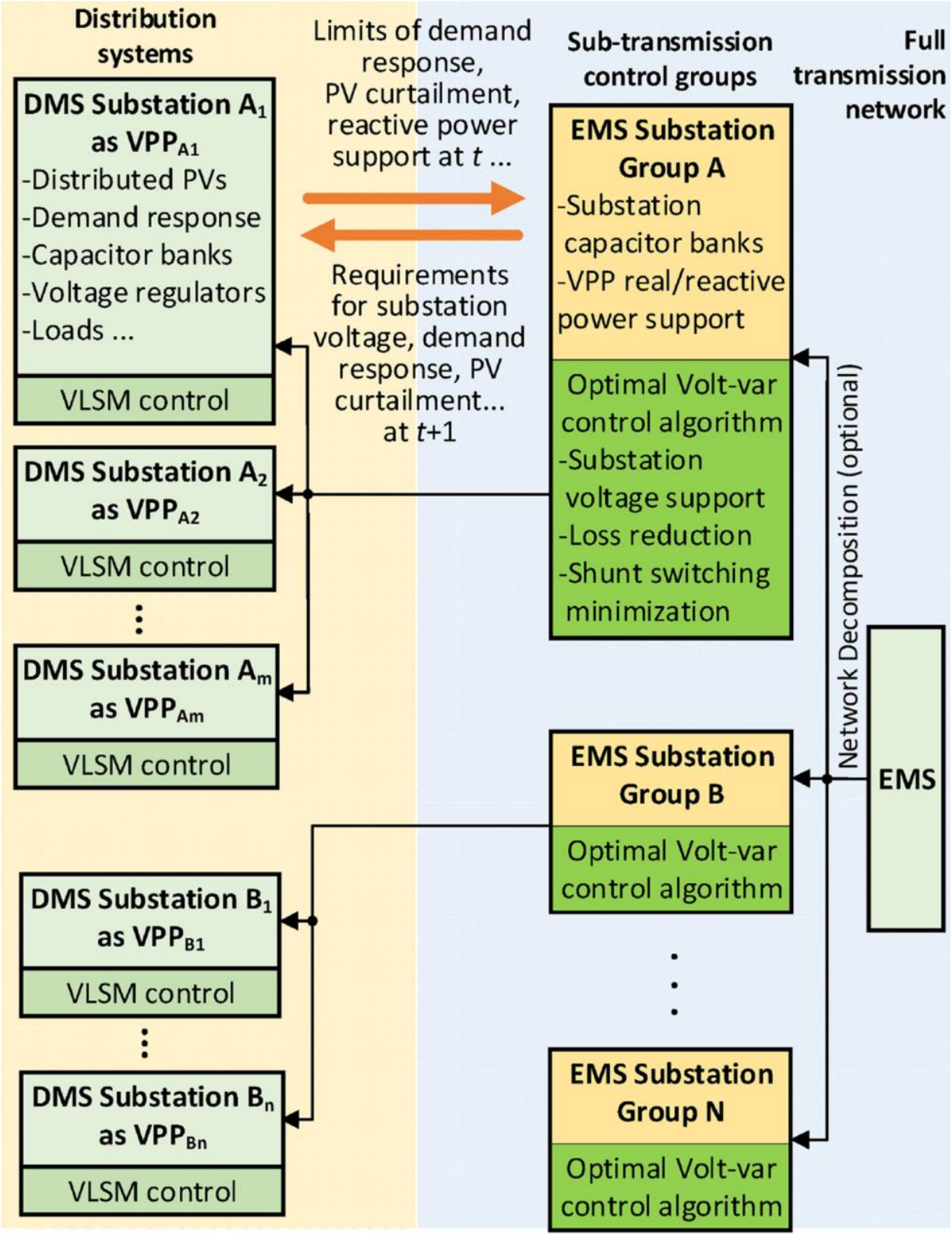}
		\vspace{-0.2cm}
		\caption{CReST-VCT control architecture \cite{Xinda_1}.}
		\label{fig:CReST_Operation}
	\end{figure}
	\vspace{-0.15cm}
	\subsection{Control Architecture}
	\vspace{-0.15cm}
	The control architecture of CReST-VCT is shown in Fig. \ref{fig:CReST_Operation}. The CReST-VCT algorithm performs coordinated optimizations between the EMS and DMS at regular intervals. The frequency of operation for the algorithm is flexible, but it is optimal if the chosen interval is consistent with the highest data resolution of load and DER. In \cite{Xinda_1}, a 5-minute interval is chosen considering the available 5-min solar data.
	
	At the beginning of each control interval, operation limits of the distribution systems are reported to the EMS residing at the sub-transmission control center. CReST-VCT, which is integrated in the EMS, runs an optimal volt-var-control (VVC) algorithm that considers all sub-transmission operational constraints to determine the operating points for var resources in the sub-transmission system as well as send the operating points for the associated distribution systems back to their DMS at the distribution control centers.
	
	At the DMS, a voltage-load sensitivity matrix (VLSM)-based VVC algorithm was developed to achieve the optimal dispatch of DERs at the distribution level \cite{Xiangqi_1}. As shown in Fig. \ref{fig:CReST_Operation}, after receiving the operation points from the energy management system (EMS), the VLSM-based VVC will dispatch DER resources, e.g. smart thermostats, energy storage devices and PV inverter units, to follow the operation point sent to it while satisfying distribution system operational requirements and constraints. The VLSM-based VCC algorithm also will calculate the upper/lower limits of the real/reactive powers of the distribution feeders and forecast the next real power operating point for the next time step. That information is sent back to EMS to be used in the next control interval.

	\vspace{-0.15cm}
	\subsection{Required communication and time flow}
	\vspace{-0.1cm}
	Preliminary tests of CReST-VCT indicate the following time flow for one operating interval:
	\begin{itemize}[leftmargin=*]
		\item The CReST-VCT algorithm at a EMS takes about 45 seconds when applying for a large-scale sub-transmission system with 3246 buses \cite{Xinda_1, Malini_1}.
		\item About 15 seconds are expected for sending the request on substation voltage, PV curtailment, and demand response from EMS to DMS. Because an EMS might communicate with multiple DMSs, the communication may be staggered within those 15 seconds.
		\item At a DMS, solving the VLSM-based volt-var control algorithm takes about 30 seconds. 
		\item The bidirectional messages about real-time dispatch and supporting reactive power capacity between DMS and downstream DERs may take up to 120 seconds over various elements of communication infrastructure.
	\end{itemize}
	In addition, the exchange of the limits of demand response, PV curtailment, and reactive power support from DMSs to EMS for the next interval may take about 60 seconds.

\section{Proposed CReST-VCT Design Framework at EMS and DMS}
	This section proposes an implementation framework of the CReST-VCT in an operational EMS and DMS systems and their interfaces. While only one approach is described, key considerations that may drive design choices are presented throughout to help tailor it to a particular operational environment.

	\subsection{CReST-VCT system structure}
	
	\subsubsection{EMS Subsystem}
	CReST-VCT would reside as one of the advanced applications in an EMS. It would have access to the data monitored by supervisory control and data acquisition (SCADA) and be able to store data in a historian or database. CReST-VCT would also require the state of the system from the state estimator application.
	
	The current implementation of CReST-VCT acts as a solver for optimal power flow (OPF) using the General Algebraic Modeling System (GAMS) modeling language and commercial solver KNITRO. Many EMS systems are also equipped with OPF tools. CReST-VCT can either leverage the built-in tools or maintain stand-alone functionalities, so as to easily port between various vendors.
	
	It is not feasible to specify the application program interfaces (APIs) for the CReST-VCT advanced application to be included within the EMS system, because this is usually proprietary information of the EMS system providers. An option to minimize API issues is to containerize the CReST-VCT advanced application using a platform like Docker. This would provide a common implementation that could be ported to a variety of different EMS systems. However, the EMS integration would still require translation from the EMS’s unique APIs to those established by the container. Though CReST-VCT is not available as a container at this writing, that can be implemented if desired.

	\subsubsection{DMS Subsystem}
	Just as in an EMS, the VLSM-based VVC algorithm at the distribution side of CReST-VCT resides in the DMS. As shown in Fig. \ref{fig:CReST_Operation}, this application takes into account requirements from CReST-VCT and allocates these requirements to its distributed resources so that the demand-response cost and the overall voltage deviation are minimized.

	The CReST-VCT component will require telemetry data for DER and distribution assets through the SCADA system. It is assumed the telemetry data is available within the DMS and the CReST-VCT component will interface using an appropriate data access API. To issue new operating points to the distribution assets in accordance with the CReST-VCT output, the DMS may communicate directly to DERs or may interface through a distributed energy resource management system (DERMS). The protocols used for these interfaces to these devices are discussed in Section IV. 
	
	Similar to the EMS integration, the details of the internal interfaces between the CReST-VCT and the DMS system are expected to be dependent on the vendor-specific DMS software and are not specified here. It is assumed that the CReST-VCT software will need to conform to the appropriate vendor-specific API. Again, this does not preclude the existing CReST-VCT software implementation from being implemented in a container as a stand-alone element or the DMS from implementing appropriate interfaces to the container.

	\begin{figure}[ht]
		\centering
		\includegraphics[width = 1\columnwidth] {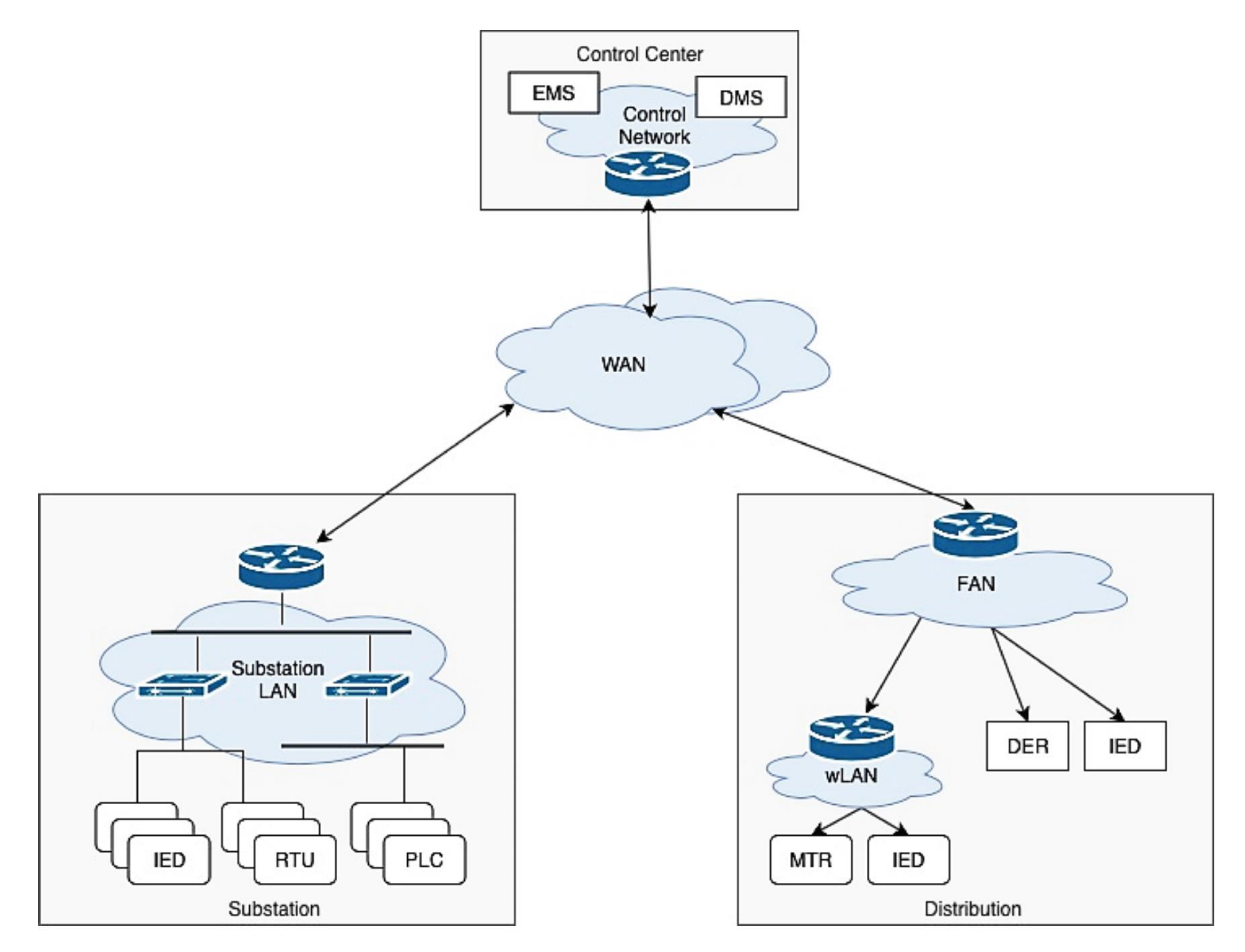}	
		\vspace{-0.5cm}
		\caption{Communication Networks Supporting CReST-VCT.}
		\label{fig:Communication}
	\end{figure}

	\vspace{-0.5cm}
	\subsection{CReST-VCT system interfaces}
	In this section, the communication interfaces between the components used by the CReST-VCT solution are identified. These communication flows cross several different communication
	network domains in the utility operational environment and use several different protocols. The diverse communication network landscape was discussed in \cite{Ogle_1}. For reference, the high-level network  diagram is shown in Fig. \ref{fig:Communication}.

	\subsubsection{EMS-DMS interface}
	Connection from EMS to DMS is often point-to-point, dedicated connection. The messages that would be passed through this channel include the reactive power required to maintain voltage
	stability in the sub-transmission network, substation voltage, aggregated PV curtailment, and 	demand-response requirement. No generally accepted standard protocol was identified for this
	interface in our research. This may be a sign of the relatively new nature of coordination between EMS and DMS. Because both EMS and DMS systems use Distributed Network Protocol 3 (DNP3) to communicate with power system components, our assumption for this
	paper is that the EMS-to-DMS interface to support CReST-VCT would be implemented in DNP3.
	
	\subsubsection{DMS to inverters}
	Connection from DMS to DERs is an area of rapid evolution. Standards are being updated or 	developed to take into consideration the variability in the many DERs systems from different vendors. California Rule 21 and Hawaii Rule 14 have mandated communication capabilities and availability of advanced functions in solar inverters with remote controllability \cite{CA_Rule_21, PG_E_1}. These developments are timely, as the penetration of PVs has been increasing, especially in the states of California and Hawaii. The IEEE Standard 2030.5 defines open standards for communications and data format for DERs \cite{IEEE2018b}. IEEE Standard 1547-2018 specifies standard choices for communication at the DER as well as mandatory advanced capabilities \cite{IEEE2018a}. Such standards support realization of requirements such as those specified by California Rule 21. However, a variety of different protocol options are still in use today. 
	
	Utilities might face interoperability challenges for communication networks. Each utility standardizes internally, adopting the collection of standards that best suits their particular environment, and uses it to dictate procurement decisions. Therefore, the protocols used to integrate CReST-VCT into a utility’s operations may differ according to the network architecture and technology of that given utility. It is expected that the chosen set of protocols and the communication networks used will be compatible with various configurations of DERs.

\section{Performance Analysis}
	\subsection{Potential communication variation}
	CReST-VCT operation relies on a bidirectional flow of operational parameters and constraints between the T\&D sides, as mentioned in Section II.C. The performance of the system depends greatly on the latency of the communication process. The faster a message flows between the two sides, the better performance system achieves. 
	
	However, there is a significant potential variation in the performance of communication networks for CReST-VCT implementations. First, the communication latency and 	bandwidth vary with the chosen protocol and the layer of the network where the control resources of CReST-VCT, such as an inverter-based PV, are connected. For example, a DER connected at the neighborhood area network layer will have significantly more latency than a device connected at the substation local area network. Feeder devices such as regulators, capacitor banks, or DERs connected at the field area network layer or the neighborhood area network layer are subject to the widest range of possible communication network performance. Advanced metering infrastructure networks also contribute to the variation because their data rates can range from 10 kbps to 1.2 Mbps. Second, the specific geographic location with reference to network infrastructure, the surrounding topology, and network utilization also introduce variations in performance in a communication network such as a wireless network \cite{Gember_1}.

	\vspace{-0.1cm}
	\subsection{Effects of communication delays}
	As described above, CReST-VCT operation spans multiple tiers of a typical utility’s communication network. Different communication system technologies may be used in each, with its own performance and reliability characteristics. 
	
	Based on control architecture in Section II.B and \cite{Xinda_1},  if solving the optimization problems at both EMS and DMS and sending data through the communication network takes $x$ minutes, the system is only expected to achieve the minimum losses while maintaining all operational constraints in the remaining (5 - $x$) minutes. As a result, the CreST-VCT solution will be more meaningful in minimizing losses and satisfying all the operational constraints over a longer period if $x$ is significantly small compared to 5 minutes. 

	\subsection{Scalability}
	CReST-VCT assumes that the DMS is fully capable of establishing two-way communication with all its control resources. Therefore, centralized control in a distribution system will reach a scalability limit as the computational effort and communicational message flow are constrained by the 5-minute 	interval in CReST-VCT input to guarantee a desired level of system performance. Therefore, the distribution side might devise decentralized and distributed approaches to control DERs locally with significantly less stringent or no communication requirements, respectively. However, these approaches only consider local state measurements rather than comprehensive system network constraints. Therefore, it cannot match the CReST-VCT solution in terms of minimizing losses and guaranteeing all operational constraints, such as voltage stabilization. Another possible solution is to make a trade-off between the nominal 5 minutes specified in CReST-VCT
	design and the scale of hosted inverter-based PVs. With an extended interval such as 10 minutes or 15 minutes instead of 5 minutes, fewer limitations are put on communication and the
	number of DERs. 
	
	The CReST-VCT tool also provides a mechanism to parallelize the algorithm computation by dividing the system into substation groups. However, each of these groups is expected to be
	synchronized to the same 5-minute operational interval to make sure the DER dispatch and supporting field measurements are coordinated across the grid. This means that the DER dispatch and field measurements will be executed in parallel. This could pose scalability issues for the communication network infrastructure.
	
	SCADA systems at the substation level operate at the typical poll cycle of one to several seconds for all points. Because these points are maintained continuously, it is assumed that
	communication networks will be scaled appropriately for normal operations, independently of CReST-VCT. Connection to devices outside the substation to take advantage of CReST-VCT
	optimization capabilities may drive new connections to the SCADA system. Further, CReST-VCT increases the potential hosting capacity of DERs that might otherwise be curtailed,
	increasing the possible number of DERs a given feeder can support. The potential optimizations with a 5-minute interval could again provide incentive to connect more devices and interact
	with those devices more often.

	\subsection{Security considerations}
	CReST-VCT has features such as remote control and remote access to DERs, and those DERs are equipped with digital communication and control interfaces that pose as cyberattack surfaces. Also, larger numbers of third-party devices associated with DER deployment can also increase cybersecurity risk if the devices are not sufficiently secure. Communication protocols used for CReST-VCT are vulnerable to attacks by attempts to intercept, modify, and/or corrupt the control signal packets. The U.S. National Electric Sector Cybersecurity Organization Resource (NESCOR) Technical Working Group 1 describes some realistic cyber threats that concerning the DER domain \cite{NESCOR_1}. Each of the Open Systems Interconnection (OSI) communication layers in the DER domain is vulnerable to potential cyberattacks. Cybersecurity should be considered with any type of EMS/DMS applications that are related to DER devices and their respective communication channels. The current cybersecurity postures of IEEE 2030.5, IEC 61850, and IEEE 1815 are discussed below.

\section{Implementation of the Distribution Side of CReST-VCT in GridAPPS-D}
	This section describes at a high level an example of how to integrate the distribution side of CReST-VCT in the open-source and application development platform GridAPPS-D. 

	\subsection{Overview about GridAPPS-D}
	As mentioned in the Introduction about the structure and characteristics of future power systems, it is important to enable a mature, efficient, and validated procedure for the development of future grid support functions for utility operators. Therefore, GridAPPS-D was introduced in \cite{GridAPPS-D_1}, and it enables distribution system application developers including commercial vendors, utilities, and research institutions to develop innovative applications related to planning and operation:
	\begin{itemize}[leftmargin=*]
		\item Researchers can use GridAPPS-D as a standardized platform to develop and test new advanced functionalities.
		\item Application developers can use GridAPPS-D to streamline a development of an application for a full range of distribution systems.
		\item System vendors can adopt the architecture and interface definitions of GridAPPS-D to support portable applications.
		\item As GridAPPS-D-compliant products enter the marketplace, utilities will be able to integrate more advanced tools that meet their specific requirements with significantly reduced cost and time into their operations.
	\end{itemize}
	In general, GridAPPS-D™ platform aims to shift utilities away from a closed, vertically integrated architecture to collaborating in an open, layered, and standards-based architecture.

	\subsection{Integrated design view for CReST-VCT in GridAPPS-D}
	\begin{figure}[t!]
		\centering
		\includegraphics[width = 0.98\columnwidth] {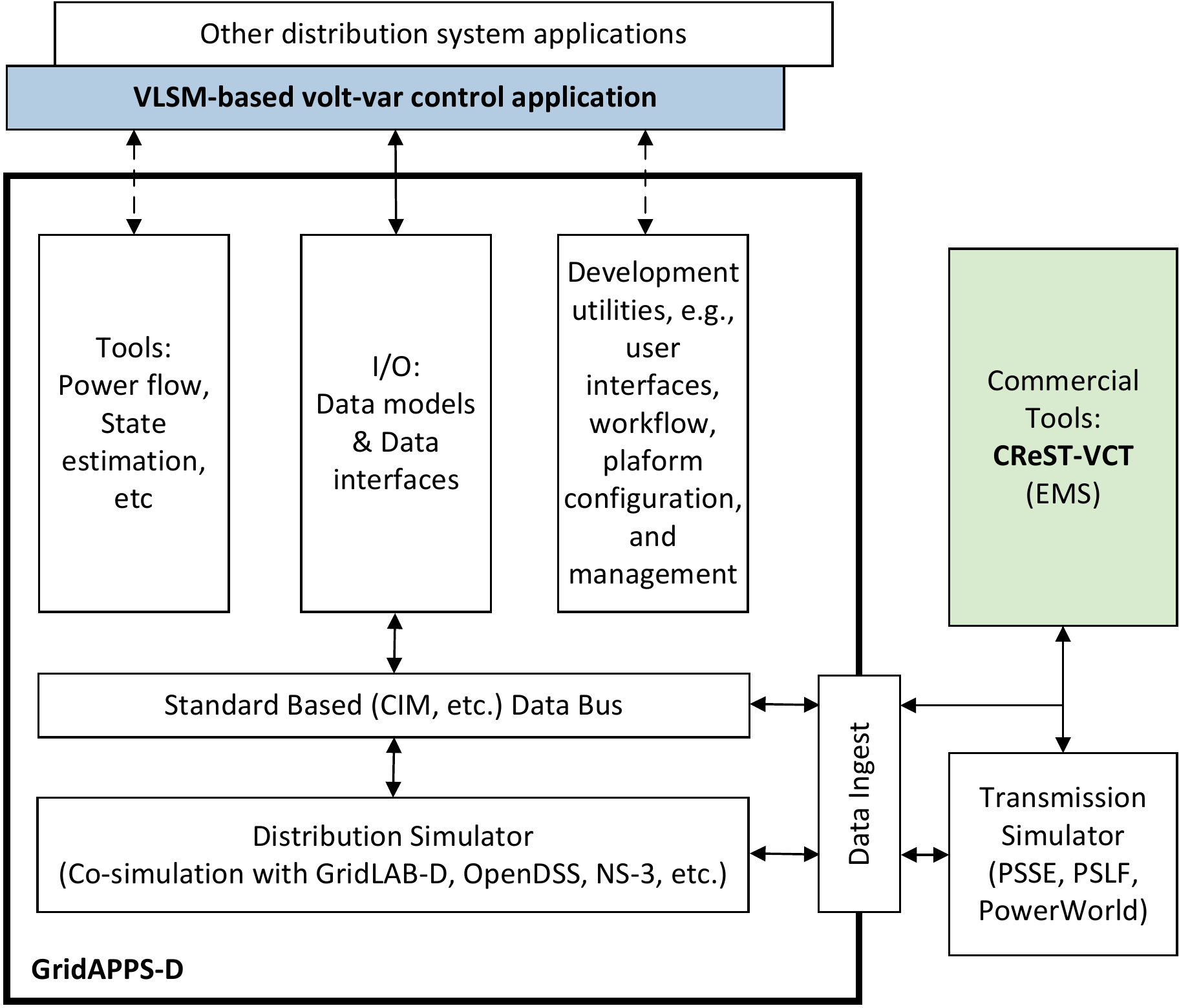}	
		\caption{The integrated design of CReST-VCT in GridAPPS-D.}
		\label{fig:CReST-VCT_GridAPPSD}
	\end{figure}
	
	As mentioned in Section 4.1, the key components of the CReST-VCT is a VLSM-based VVC algorithm installed at the DMS and a CReST-VCT algorithm installed at EMS. 

	For the integration of the co-simulation CReST-VCT in GridAPPS-D, the VLSM-based algorithm will be integrated as a new distribution system application, among other existing applications. It is a high-level application that iteratively receives the operating point determined from the EMS through the Data Ingest Interface, Standard-based Data Bus, and I/O Interface components inside GridAPPS-D. The VLSM-based algorithm then determines the optimal dispatch for the DER resources, e.g. smart thermostats, energy storage devices, and PV inverter units in the distribution model for each 5-minute period. This dispatch information is delivered to the distribution simulator, either in GridLAB-D or OpenDSS, via the I/O interface and Standard-based Data Bus components. 

	On the other hand, the CReST-VCT algorithm at the EMS is considered as an external tool that constantly coordinates with the VLSM-based VVC application by exchanging bidirectional data through the Data Ingest component of the GridAPPS-D. In addition, the transmission simulators such as PSSE, PSLF, and PowerWorld can also link and exchange information to the distribution system simulators such as GridLAB-D and OpenDSS to replicate the physical connection in practice.

\section{Conclusion}
		\vspace{-0.01cm}
	This paper presents a system integration framework for integrating the CReST-VCT and other advanced grid support function tools into the real-time operational environment of EMS and DMS. The framework serves as a guideline on designing system structure and interfaces that is consistent with the active and data driven characteristics of future transmission and distribution system under high penetration of DERs. Communication variation, negative effects of communication delay, scalability, and security are also discussed. To further clarify the proposed framework, the paper presents a high-level description about how CReST-VCT can be integrated in GridAPPS-D, which is an open-source and standardized platform supporting system application developers and DMS.

\section{Acknowledgment}
	This study is funded by the U.S. Department of Energy Solar Energy Technologies Program (SETO) through The Technology Commercialization Fund (TCF). The TCF is administered by the DOE's Office of Technology Transitions.

\bibliography{QuanNguyen_References}
\bibliographystyle{IEEEtran}

\end{document}